\documentclass{emulateapj}

\def\hub{\ifmmode H_\circ\else H$_\circ$\fi}
\def\fe{[Fe/H]}
\def\xone{$x_1$}
\def\xtwo{$x_2$}
\def\rquarter{$R^{1/4}$}

\shorttitle{Bar Globular Clusters in M31}
\shortauthors{Morrison et al.}

\usepackage{color}
\usepackage{afterpage}

\begin{document}

\title{Star Clusters in M31:  Old Clusters with Bar Kinematics}

\author{Heather Morrison}
\affil{Department of Astronomy,
Case Western Reserve University, Cleveland OH 44106-7215
\\ electronic mail: heather@vegemite.case.edu}

\author{Nelson Caldwell} 
\affil{Center for Astrophysics, 60 Garden Street, Cambridge, MA 02138, USA
\\ electronic mail: caldwell@cfa.harvard.edu}

\author{Ricardo P. Schiavon}
\affil{Gemini Observatory, 670 N. A'ohoku Place , Hilo, HI 96720, USA \\ electronic mail:
rschiavo@gemini.edu}

\author{E. Athanassoula}
\affil{LAM/OAMP, UMR6110, CNRS/Univ. de Provence, 38 rue Joliot Curie,
13388 Marseille 13, France\\electronic mail: lia@oamp.fr}

\author{Aaron J. Romanowsky}
\affil{UCO/Lick Observatory, University of California, Santa Cruz, CA
95064, USA\\electronic mail: romanow@ucolick.org}

\author{Paul Harding}
\affil{Department of Astronomy,
Case Western Reserve University, Cleveland OH 44106-7215
\\ electronic mail: paul.harding@case.edu}


\begin{abstract}

We analyze our accurate kinematical data for the old clusters in the
inner regions of M31.  These velocities are based on high S/N
Hectospec data (Caldwell et al 2010). The data are well suited for
analysis of M31's inner regions because we took particular care to
correct for contamination by unresolved field stars from the disk and bulge in
the fibers. The metal poor clusters show kinematics which are
compatible with a pressure-supported spheroid. The kinematics of
metal-rich clusters, however, argue for a disk population. In
particular the innermost region (inside 2 kpc) shows the kinematics of
the \xtwo\ family of bar periodic orbits, arguing for the existence of an
inner Lindblad resonance in M31. 

\end{abstract}


\section{Introduction}

Globular clusters can provide simultaneous estimates of velocity,
metallicity and age: a powerful trio with which to study the history
of a galaxy. They are particularly helpful to complement integrated
light studies, which average over all stellar populations along a line
of sight. In this paper we discuss the kinematics of old clusters
projected on the inner 10 kpc of M31.  Roughly one third of our sample
of over 300 old clusters in M31 (presented in Caldwell et al.\ 2009 and
2010, Papers 1 \& 2 hereafter) are located within 3 kpc of its
center. Because of our careful treatment of the effect of field star
contamination from the bright bulge and inner disk region in our
fibers, our dataset is particularly well suited for study of the
central regions of M31.

Early work on bulge kinematics \citep{defis} showed that bulges
resemble low-luminosity ellipticals in being kinematically hot with a
high degree of rotational support ($V/\sigma \sim 1$). Later studies,
however, showed that bulges are more complex and that an important
distinction must be made between classical \rquarter\ bulges -- which
are kinematically hot and formed rapidly from mergers and collapses --
and bulges formed via secular evolution of disks, which have a lower
Sersic index (Kormendy \& Kennicutt 2004). In this second category
Athanassoula (
2005) distinguished the boxy/peanut bulges -- which are
parts of bars seen edge-on -- and the disk-like bulges, which have a
disk shape.  Boxy/peanut bulges can be distinguished in near-edge-on
galaxies from photometry or via kinematics \citep[eg][]{konrad95,martinlia99}.

Evidence from isophotal twists and kinematics was used to argue that
M31 might have a triaxial bulge or a bar (Lindblad 1956, Stark 1977,
Stark \& Binney 1994). More recently, Athanassoula \& Beaton (2006) and
Beaton et al. (2007), using deep 2MASS observations, considerably
strengthened the case for a bar and suggested that M31 also has a
centrally concentrated classical bulge, which dominates the light in
the inner 200 pc.  We note here that this is a considerably smaller
and less dominant classical bulge than the one suggested by previous
authors: \citet{devauc58} found an effective radius of 3.5 kpc, and
\citet{walterbos} derived an effective radius of 2 kpc and found that
the bulge contributed 40\% of the light of the galaxy.
Our globular cluster kinematical data allow us to further explore this
shift in our view of M31's bulge, since we have [Fe/H] and velocity
measurements for 98 old clusters projected within 3 kpc of M31's
center.

We assume a distance of 770 kpc throughout \citep{freedman} and a PA
of 37.7 degrees.  The XY coordinate system we use in this paper has
units of kpc, with positive X along the major axis towards the NE.

\section{Cluster Kinematics}

Paper 2 presented [Fe/H], age and velocity measurements based on high
S/N Hectospec \citep{fabricant} spectra (a median S/N of 75 per $\AA$)
for over 300 M31 clusters with ages greater than 6 Gyr. (In fact the
great majority of these clusters have ages greater than 10 Gyr.) Here
we discuss the old clusters from this paper which are within 2 kpc of
M31's major axis.  Repeat Hectospec observations showed a median
velocity error of 6 km~s$^{-1}$.  Our study contains 17 entirely new
cluster velocities and is the first fiber study to use offset
exposures near each cluster in the bright inner regions to correct for
the contamination from field stars there. \citet{paper2} showed that
ignoring this effect can lead to velocity errors of more than 100
km~s$^{-1}$. In the small number of cases where our velocities
differed significantly from the Hectochelle velocities of Strader and
Caldwell (2010, in preparation), we have used the more accurate
Hectochelle data.  Our [Fe/H] values are in good agreement with the
recent results of \citet{beasley05} and \citet{colucci} and the HST
color-magnitude derived values.  We found that the old cluster
metallicity distribution was neither unimodal nor simply bimodal,
showing a median [Fe/H] around --1.0 and possible peaks at [Fe/H] =
--0.3, --0.8 and --1.4.

Previous work on M31 globulars suggested a larger systemic rotation
for the metal-richer clusters \citep[eg][]{huchra,barmby}.  Since all
but one of the clusters with \fe\ $ > -0.4$ are projected on the inner
disk (less than 1.5 kpc from the major axis) we first explore connections
between the metal-rich clusters and M31's disk. In the Milky Way, the
metal-rich globular clusters are likely associated with its
inner disk: \citet{zinn85} connected these clusters with the thick
disk, \citet{dante96} with the bulge. Since more recent work has shown
that the Milky Way bulge is dominated by a bar \citep{weiland,
binney97}, the most metal-rich globulars in the Milky Way are all
likely to be connected to its disk in some way: either as a bar distortion
or a thicker component.

\begin{figure}[!h]
\includegraphics[scale=.45]{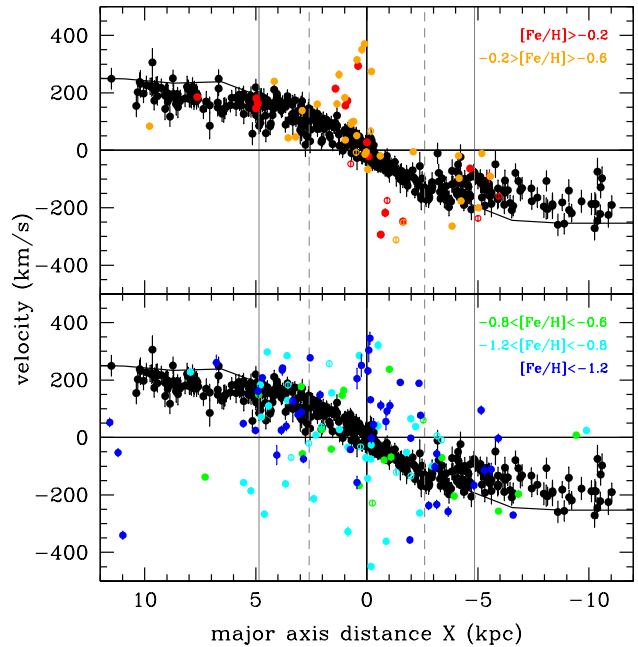} 

\caption{Velocity of clusters projected less than 2 kpc from the major
  axis. The upper panel shows the metal-rich clusters (\fe$>$--0.6)
  and the lower panel the more metal-poor clusters which dominate
  M31's old clusters.  In both panels, black symbols
  show the {\it mean} velocity of stars at that position, integrated along
  the line of sight. Open symbols denote ages less than 10 Gyr, closed
  symbols greater than 10 Gyr (note that we are unable to measure ages
  for clusters with \fe\ less than --1.0, and we use closed symbols
  for these clusters). The vertical grey lines show the end of the
  thick bar (dashed lines) and the thin bar (solid), from \citet{beaton} and
  \citet{liabeaton}, and the solid black line is the rotation curve
  from \citet{kentrotcurve}.  Note that measurements of the
  dimensions of the thin and particularly the thick bar are
  approximate only.
\label{hilofekin} }
\end{figure}

To compare our old clusters with M31's disk, we use the
fibers which were designed to measure the contamination from M31's
disk and bulge: each fiber's measured velocity will be the {\it mean}
velocity of all the stars along that line of sight. Figure
\ref{hilofekin} compares cluster velocities (with different colors for
clusters with different metallicities: red for most metal rich
through blue for the most metal poor) with these mean velocity
estimates, shown in black. The top panel shows clusters with
\fe\ $>-0.6$; the bottom panel shows more metal poor clusters.
The difference between their kinematic behavior is stunning.

Metal poor clusters (lower panel) show little sign of rotation and occupy
the four quadrants of the plot similarly. On the other hand, the
metal-rich clusters (upper panel) show a distinct and quite cold
kinematical signature. There are almost no clusters in the forbidden
quadrants (occupancy here corresponds to rotation in the opposite
direction to the disk) and most of those more than 2 kpc from the
center (ie $|X|>2$) have velocities which closely follow the disk
velocity at that position. 
However, in the inner 2 kpc the signature differs from the usual one
for a disk composed of stars on near-circular orbits.  Although all
except one cluster occupy the same quadrant as the disk, thus respecting the
same direction of rotation, their velocities can deviate from the local
mean velocity of the integrated light by up to 350 km~s$^{-1}$
(recall that M31's rotation
velocity is 250 km~s$^{-1}$). We note that very high velocities are
also observed in the HI gas in this region \citep{brinkshane}. 
In the following section we will
describe expectations for the kinematics of thin disk, bar and
classical bulge objects, and show that this signature is expected for
bar orbits.

\subsection{Kinematics: expectation from disk, bar and bulge}

Thin disks in galaxies have ``cold'' kinematics dominated by rotation
and show a low velocity dispersion. We showed in Paper 1 (see Figure 13)
that the young M31 clusters (with ages less than 2 Gyr) have such
kinematics: the young clusters all follow the same narrow
locus in position vs velocity. We also showed (see Figure 12)
the mean velocity field across the face of the disk, obtained
from our ``sky'' fibers. The mean velocity changes smoothly and slowly
as we look from the receding side of the disk through the center to
the approaching side, as expected for a thin disk.

It is particularly simple to follow the kinematic signature of a cold,
thin disk by examining velocities of objects seen close to the
major axis. In a galaxy close to edge-on such as M31, such a star in a
circular orbit will have all its velocity in the line of sight, giving
a clean measure of $V_\phi$, the azimuthal component, from the
line-of-sight velocity. For disk stars observed at larger distances
from the major axis, less of their azimuthal velocity will be
projected onto the line of sight and so the change in mean velocity
from one side of the disk to the other will be smaller.

\begin{figure}[!h]
\includegraphics[scale=.3]{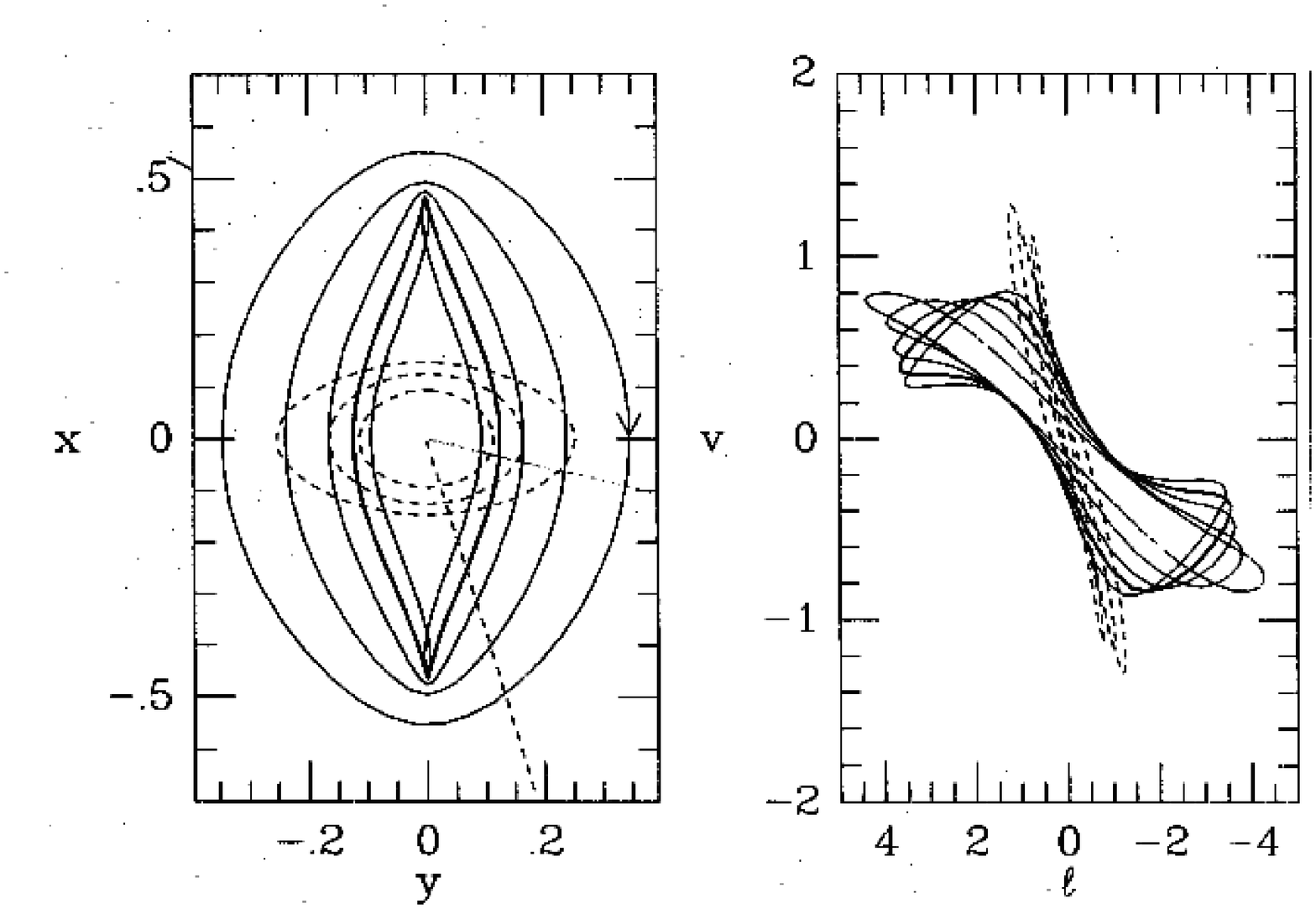}
\caption{Illustration of regions occupied both spatially and
kinematically by \xone\ and \xtwo\ orbits in a barred potential, from
\citet{binney91}. Left panel shows their spatial location in a face-on
view, while the right panel shows longitude-velocity plots. \xone\ orbits are aligned along the bar major axis
and shown with solid lines, while \xtwo\ orbits align perpendicular to
the bar major axis and are shown with dotted lines.  Note that in this
example, the \xtwo\ orbits can reach velocities significantly higher
than the circular velocity, which is v=1 in this model.
\label{x1x2orbits}}
\end{figure}

Most orbits in bars follow the two main families of closed periodic orbits
\citep{binneytremaine}: the \xone\ orbits, which are aligned along the
long axis of the bar (close to the major axis in M31, see Beaton et al
(2007)) and \xtwo\ orbits which are aligned along its short axis (close to M31's minor axis).
 For \xtwo\ orbits, velocities can reach very high values
close to the center of the galaxy.  
This is due to the fact that they are
observed near-end-on, so that the line-of-sight component is nearly
along the orbit at its pericenter \citep{martinlia99}.
Figure \ref{x1x2orbits}
\citep[from][]{binney91} illustrates the spatial and velocity
signatures of \xone\ and \xtwo\ orbits. It can be seen that in this
example, the \xtwo\ orbits reach velocities much higher than the circular
velocity. \citep[A similar position-velocity diagram for a M31-like
  system can be seen in the middle top panel of Figure 11
  of][]{liabeaton}.

Lastly, we would expect any classical bulge component to show
$V/\sigma \sim 1$: some rotational support but a roughly equivalent
amount of random motion. \citet{kent89} fitted the M31 bulge using an
oblate rotator model with major-axis velocity of around 90 km~s$^{-1}$
and velocity dispersion of 130 km~s$^{-1}$ at 1.5 kpc from the center.

\section{Discussion}

We saw in Figure \ref{hilofekin} that the kinematics of old M31
clusters with \fe\ $> -0.6$ in its innermost region show the
distinctive behavior of objects on \xtwo\ orbits in M31's bar.  This
is in very good agreement with orbital structure in bars since the
\xtwo\ orbits are always confined to the innermost regions, in the
region interior to the inner Lindblad resonance (ILR).  Most of the
rest of the metal-rich clusters have orbits consistent with disk
objects. 

We see little or no indication in the kinematics in the upper panel of
Figure \ref{hilofekin} for a kinematically hot population such as the
classical bulge of \citet{kent89}. However, we note again that the
classical bulge identified by \citet{beaton} was quite small, only
dominating the inner 200 pc. We have only one cluster within 200 pc of
M31's center in our sample, so we cannot probe the kinematics
of this region in M31.
Only in the lower panel, with the more metal-poor clusters, do we see
a signature like that of a kinematically hot classical bulge: there
are roughly equal numbers of clusters in each quadrant, and we see
that the velocity dispersion rises sharply close to the center, as we
would expect for a centrally concentrated classical bulge. However, as
we shall show below, the starlight in this region is dominated by
old metal-rich stars of near solar abundance, so these metal-poor
clusters are not tracing the dominant component here. 

To summarize: we see strong evidence from the kinematics of the
metal-rich old clusters ([Fe/H]$>$--0.6) for both disk and bar
kinematics. A number of the clusters within 2 kpc of the center of M31
show the kinematic signature of \xtwo\ orbits in a barred
potential. The rest of these clusters (plus the other metal-rich
clusters within 2 kpc of the major axis) show the cold kinematics of
the disk. These kinematics strongly confirm the result of
\citet{beaton} and \citet{liabeaton} that M31 has a bar whose inner
parts constitute the boxy bulge which dominates its light in the inner
few kpc. To our knowledge, this is the first clear detection of
globular clusters with bar kinematics in any galaxy. However, there is
one massive cluster (the Arches cluster) in the Milky Way which
has a large space velocity (232 km~s$^{-1}$) and is currently at a
projected distance of only 26 pc from the galactic center
\citep{stolte08}. These authors note that the cluster could be on a
transitional trajectory between \xone\ and \xtwo\ orbits in the Milky
Way's barred potential, and may have been formed in a starburst
triggered when a massive molecular cloud ``collided on the boundary
between \xone\ and \xtwo\ orbits in the inner bar''.  

\subsection{Relations between clusters and bulge/disk field stars}

We now consider the relationship between field stars and
globular clusters in the inner regions of M31.
\citet{trager00} studied the integrated light of M31's bulge, in a
circular aperture of diameter 250 pc. They found a mean metallicity of
+0.2 dex, and a mean age of around 6 Gyr. More recently,
\citet{saglia} have made a detailed study of the M31 bulge region
using a number of long-slit exposures with the HET. They find a mean
metallicity around solar, and an age of around 12 Gyr in the inner 1-2
kpc. (Note that they do see a metallicity gradient, reaching up to
[Z/H]=+0.4, over the inner 200 pc, the region dominated by the
classical bulge.) 

 \citet{ata05} used HST/WFPC2 observations to produce a
color-magnitude diagram for M31's bulge at 1.6 kpc from its center,
and inferred a metallicity distribution which peaked near solar.
\citet{olsen} summarized near IR color-magnitude diagrams from high
spatial resolution studies of M31 to find
that the stellar population in the inner few kpc was dominated by old,
nearly solar-metallicity stars. Interestingly, by comparing fields in
the bulge with an inner disk field, they found no evidence for an age
difference between bulge and disk. This is unsurprising if M31's bulge
is dominated by a bar, since bar stars are merely inner disk stars
which have become part of the bar pattern.

The mean metallicity of the integrated light from field stars thus exceeds
the mean metallicity of the globulars in the inner few kpc; it is
closer to the mean of those with [Fe/H]$>$--0.6, which show either 
disk or bar kinematics. 
(It has been suggested before that globular
clusters are formed less efficiently in metal-rich populations:
\citet{strader05} calculate that the efficiencies differ by more than a factor of
10 in the Milky Way, by comparing metal-rich globular clusters to the
bulge luminosity and metal-poor numbers to the halo luminosity. This
number will not be changed radically if we substitute the thick disk
luminosity for the bulge luminosity in this calculation.)

Thus a simple picture can explain the existence of the metal-rich
globular clusters in M31: they merely participated in the early
formation of the inner disk and  the onset of the bar instability.

\section{Summary}

We have discussed accurate kinematical data for old M31 clusters in
its inner regions within 2 kpc of its major axis. The majority of the
metal-rich clusters (those with [Fe/H] greater than --0.6) show disk
kinematics, and many
of the clusters within the innermost bar region have the 
signature of the \xtwo\ family. This clearly shows the existence
of an ILR and, to our knowledge, this is the first time it has been
clearly shown using stellar kinematics. In the only other known
example,  \citet{teuben} showed this using gas
  kinematics in the strongly barred galaxy NGC 1365. Our result
also gives an estimate of the ILR location, which
provides useful constraints for future dynamical studies of M31 since
it could be used to set limits to the bar pattern speed. These
metal-rich clusters share the
population properties (metallicity and age) of the integrated light in
the inner few kpc, which has been studied both via spectroscopy and
via deep color-magnitude diagrams from HST and adaptive optics
imaging.  By contrast, clusters with \fe\ less than --0.6 within 2 kpc
of the major axis show little rotational support and a velocity
dispersion which increases as radial distance to the center decreases.

Our data do not probe the small region (200 pc) occupied by M31's classical
bulge in the description of \citet{beaton}, so we cannot comment on
its kinematics. However, we caution against simply interpreting a high
velocity dispersion in M31's inner few kpc as a bulge velocity
dispersion and then using it to constrain M31's black hole mass
\citep[as done most recently by ][]{saglia}: the contribution of the
bar, which dominates the light there, needs to be assessed.

\acknowledgements

HLM thanks the NSF for support under grant AST-0607518, AJR for grants
AST-0808099 and AST-0909237, and EA the ANR for ANR-06-BLAN-0172.
RPS is supported by Gemini Observatory, which is operated by AURA,
Inc, on behalf of the international Gemini partnership of Argentina,
Australia, Brazil, Canada, Chile, the United Kingdom and the United
States of America. We also thank John Wiley and Sons for permission to
reproduce Figure 2.

\end{document}